\documentclass[reprint,aip,aps]{revtex4-1}

\usepackage{graphicx}
\usepackage{amsmath,amsfonts}

\usepackage{amssymb}
\usepackage{bm}

\begin{document}

\title{Time-domain simulation of ultrasound propagation in a tissue-like medium based on the resolution of the nonlinear acoustic constitutive relations}

\author{No\'e Jim\'enez}
\affiliation{Instituto de Investigaci\'on para la Gesti\'on Integrada de Zonas Costeras, Universitat Polit\`ecnica de Val\`encia, Paranimf 1, 46730 Grao de Gandia, Spain}

\author{Francisco Camarena}
\affiliation{Instituto de Investigaci\'on para la Gesti\'on Integrada de Zonas Costeras, Universitat Polit\`ecnica de Val\`encia, Paranimf 1, 46730 Grao de Gandia, Spain}

\author{Javier Redondo}
\affiliation{Instituto de Investigaci\'on para la Gesti\'on Integrada de Zonas Costeras, Universitat Polit\`ecnica de Val\`encia, Paranimf 1, 46730 Grao de Gandia, Spain}

\author{V\'ictor S\'anchez-Morcillo}
\affiliation{Instituto de Investigaci\'on para la Gesti\'on Integrada de Zonas Costeras, Universitat Polit\`ecnica de Val\`encia, Paranimf 1, 46730 Grao de Gandia, Spain}

\author{Yi Hou }
\affiliation{Department of Biomedical Engineering, Columbia University, 351 Engineering Terrace, mail code 8904, 1210 Amsterdam Avenue, New York, NY, USA}

\author{Elisa E. Konofagou}
\affiliation{Department of Biomedical Engineering, Columbia University, 351 Engineering Terrace, mail code 8904, 1210 Amsterdam Avenue, New York, NY, USA}
\altaffiliation[Also at: ]{Department of Radiology, Columbia University, 351 Engineering Terrace, mail code 8904, 1210 Amsterdam Avenue, New York, NY, USA}

\date{The present paper is a preprint, submited to J. Acoust. Soc. Am. in June 2015}

\begin{abstract} 
A time-domain numerical code based on the constitutive relations of nonlinear acoustics for simulating ultrasound propagation is presented. To model frequency power law attenuation, such as observed in biological media, multiple relaxation processes are included and relaxation parameters are fitted to both exact frequency power law attenuation and empirically measured attenuation of a variety of tissues that does not fit an exact power law. A computational technique based on artificial relaxation is included to correct the non-negligible numerical dispersion of the numerical method and to improve stability when shock waves are present. This technique avoids the use of high order finite difference schemes, leading to fast calculations. The numerical code is especially suitable to study high intensity and focused axisymmetric acoustic beams in tissue-like medium, as it is based on the full constitutive relations that overcomes the limitations of the parabolic approximations, while some specific effects not contemplated by the Westervelt equation can be also studied. The accuracy of the method is discussed by comparing the proposed simulation solutions to one-dimensional analytical ones, to $k$-space numerical solutions and also to experimental data from a focused beam propagating in a frequency power law attenuation media.

\end{abstract}


\maketitle

\section{Introduction}
	The significant development of ultrasound technology in the medical field in recent years has led to the need for simulation tools increasingly realistic. Effects like absorption in biological media, nonlinear propagation, heterogeneities, strong focusing, streaming, resonances, multiple scattering or the presence of discontinuities due to tissue layers or rigid boundaries have to be taken into consideration. The most general approach for ultrasound simulation is to directly solve the constitutive relations of the nonlinear acoustics. It also allows the explicit calculation of the particle velocity, what can be used to compute important magnitudes as the vector components of the nonlinear acoustic intensity or the acoustic radiation force.
    
	The numerical resolution of the nonlinear constitutive equations in tissue-like medium supposes a difficult problem due to the large size of the region of interest in relation to the size of the acoustic wavelength and the complexity of the model. Simplifying assumptions have been needed in the past for modeling beam patterns from ultrasound transducers, as one-way parabolic approximations, most based on the Khokhlov-Zabolotskaya-Kuznetsov equation (KZK) \cite{Aanonsen1984,Lee1995,Cleveland1996,Yang2005,Khokhlova2006,Jing2007,Soneson2007,Prieur2011}. To overcome the validity limitations of the parabolic approximations, i.e. for large aperture focused sound sources or modeling sound field near the acoustic source, many one-way numerical methods have been proposed, including phenomenological approaches \cite{Christopher1991,Tavakkoli1998} with tissue attenuation \cite{Zemp2003}, or based on the one-way formulation of the Westervelt equation \cite{Varslot2005,Yuldashev2011}.
    
    Tissue inhomogeneity can be modeled in these one-way models \cite{Jing2007}, like transmission though tissue layers with refraction, but they do not take into account backscattering and multiple reflections. More realistic models, e.g. those accounting for scattering from internal tissue structures, are based on the Westervelt-type full-wave equations \cite{Hallaj1999,Hallaj2001,Pinton2009,Huijssen2010,Demi2011,Verweij2013,Jing2012}. This full-wave equation has been validated for strongly focused sources \cite{Jing2011}. However, due to the assumptions taken in the derivation of the Westervelt equation, the accuracy of this model is limited in practical situations as ($i$) the modeling of rigid boundaries where the thermo-viscous boundary layer effects are not-negligible, i.e. in general case where the particle velocity field becomes rotational, ($ii$) situations where the second order Lagrangian density of acoustical energy not vanish, i.e. near the source or in situations where plane progressive waves does not exist and the acoustic field becomes complex due to multiple scattering, reverberation or resonances, ($iii$) situations where the equilibrium-state particle velocity is not null, including the self generation of acoustic streaming. See Ref. \citet{Hamilton1998} Chap. 3 for further discussion. 
    
    The recent development of computational capacity has made possible to consider the full constitutive relations (i.e. without the assumptions discussed above). Thus, for small-amplitude acoustic waves, the linearized pressure-velocity formulation of constitutive relations in inhomogeneous media was solved by means of Finite-Differences in Time-Domain (FDTD) methods with frequency independent losses by Manry et al. \cite{Manry1996}, or using two-step MacCormack finite-differences scheme by Mast et al. \cite{Mast1997}. Also, relaxation processes can be included in finite difference methods in an efficient way in order to model tissue attenuation and dispersion \cite{Yuan1999}. Furthermore, $k$-space numerical methods have also been applied to solve the linearized first order equations in lossless inhomogeneous media \cite{Mast2001}. In order to account for soft tissue losses, the computational solution of the fractional Laplacian by $k$-space spectral methods have demonstrated to be extremely efficient due to the spatial frequency domain representation of the acoustic field \cite{Treeby2010}.

	In the case of nonlinear constitutive relations models, the evolution of the acoustic magnitudes have been simulated in time-domain by means of finite differences schemes such as Dispersion Relation Preserving method (DRP) in ideal fluids and axisymmetric domains \cite{Ginter2002}. Thermo-viscous losses in finite-differences methods have been widely used, see Sparrow et al. \cite{Sparrow1991}. In order to introduce tissue attenuation in the governing equations time-dependent fractional derivatives can be included by convolutional operators. Thus, in Ref. \citet{Liebler2004} an efficient method has been presented, but although the memory requirements can be strongly reduced compared to direct convolutions the algorithm employs up to ten auxiliary fields and a memory buffer of three time steps. Furthermore, construction of specific causal memory functions that models soft tissue attenuation and dispersion in Navier-Stokes equations is also possible \cite{Lobanova2014}, but certain time history must be stored in memory and in this case the computational domain was restricted to one dimensional propagation.

	In order to overcome those numerical limitations, recently $k$-space and pseudo-spectral numerical methods have been applied to constitutive relations in nonlinear regime to solve fractional Laplacian operators efficiently \cite{Treeby2012}. Furthermore, in the case of domains of hundreds of wavelengths, when the cumulative phase error due to numerical dispersion of standard finite-difference schemes can not be neglected, those spectral numerical methods have reported an improvement in accuracy of the numerical solution. This two factors, i.e. the negligible numerical dispersion and the efficient resolution of fractional Laplacian operators, have led the spectral methods to be widely used in practical applications. However, their main limitation is that the implementation of natural space discontinuities due to tissue layers or rigid boundary conditions leads to errors in the reconstruction of the spectral information due to the poor convergence of Fourier series at jumps, i.e. the well-known Gibbs oscillations. Preventing this kind of errors is typically achieved by filtering the spatial spectrum \cite{Jing2012}, so the theoretical spatial minimum sampling of two point per wavelength becomes larger. In addition, these errors propagate globally and affect to the accuracy all over the domain, in contrast with locally propagating errors in finite differences methods. On the other hand, taking into account the spatial discontinuity due to symmetry boundary condition, axisymmetric domains becomes not feasible by standard $k$-space methods, and full 3D domains must be employed even for axisymmetric configurations. Those errors can be prevented by means of the recently developed Fourier Continuation (FC) method \cite{Albin2012}. However, the discontinuities formed due to shock propagation are still not solved by FC methods and other additional numerical treatment must be applied for correctly describe shock formation, e.g. intensive computations by high order accurate weighted essentially non-oscillatory schemes (FC/WENO) \cite{Shahbazi2011}. Unlikely, the computational times increases by using those intensive computational techniques and the multiresolution analysis to detect discontinuities in the domain.
    
    The aim of the present work is to present a generalization of the constitutive relations of nonlinear acoustics including multiple relaxation processes in a non-convolutional formulation that allows the time-domain numerical solution by an explicit finite differences numerical scheme. Frequency power law attenuation based in relaxation have been applied in the same way than it has been applied to generalized Burgers equation \cite{Cleveland1996}, Khokhlov-Zabolotskaya-Kuznetsov (KZK) model \cite{Cleveland1996,Yang2005} and Westervelt equation \cite{Pinton2009}. The relaxation parameters have been fitted to both exact frequency power law attenuation and empirically measured attenuation of a variety of tissues that does not fit an exact power law. Two processes have been enough to model tissue attenuation with acceptable accuracy over a frequency range covering about 4 octaves, as it was demonstrated by Yang et al. \cite{Yang2005}.  A numerical technique based on artificial relaxation is included to control the non-negligible numerical dispersion of the FDTD method and improve stability when shock waves are present in the solution. The method includes backscattering and arbitrary propagation direction of finite amplitude beams, and can be  specially suitable in axisymmetric configurations where the computational resources for full 3D $k$-space methods are prohibitive. 
   
   The paper is organized as follows: in Sec.~\ref{s:model} the model equations that describes the problem are exposed, Sec.~\ref{s:numerical} describes the computational method presented in this work and in Sec.~\ref{s:validation} the method is validated comparing the numerical results with analytic solutions for linear, smooth and discontinuous nonlinear waves. In Sec.~\ref{s:result} solutions for frequency power law media are presented and compared with analytic and numerical solutions obtained by $k$-space methods as benchmark case. Furthermore, an experimental test was presented where a focused beam propagating in castor oil has been modeled. Finally, a high intensity focused source in the high nonlinear regime focusing in soft-tissue is modeled.

\section{Generalized nonlinear acoustics model for multiple relaxation media}\label{s:model}

\subsection{Full-wave modeling}\label{s:model:nonlinear}
	The principles of mass and momentum conservation lead to the main constitutive relations for nonlinear acoustic waves, which for a fluid can be expressed as \cite{Naugolnykh1998}
\begin{equation}\label{eq:continuity}
	\frac{{\partial \rho }}{{\partial t}} =  - \nabla  \cdot \left( {\rho {\bf{v}}} \right)
\end{equation}

	\noindent and
\begin{equation}\label{eq:momentum}
	\rho \left( {\frac{{\partial {\bf{v}}}}{{\partial t}} + {\bf{v}} \cdot \nabla {\bf{v}}} \right) =  - \nabla p + \eta {\nabla ^2}{\bf{v}} + \left( {\zeta  + \frac{\eta }{3}} \right)\nabla \left( {\nabla  \cdot {\bf{v}}} \right),
\end{equation}

	\noindent where $\rho$ is the total density field, $\textbf{v}$ is the particle velocity vector, $p$ is the pressure, and $\eta$ and $\zeta$ are the coefficients of shear and the bulk viscosity respectively. The acoustic waves described by this model exhibit viscous losses with quadratic power law dependence on frequency. In order to include a power law frequency dependence on the attenuation, a multiple relaxation model will be added into the time domain equations.

	The basic mechanism for energy loss in relaxing media is the appearance of a phase shift between the pressure and density fields. This behavior is commonly modeled as a time dependent relation at the fluid state equation, that for a fluid retaining the material nonlinear effects up to second order an be expressed as \cite{Naugolnykh1998,Rudenko1977}:
\begin{equation}\label{eq:st_cont}
	p = c_0^2\rho ' + \frac{{c_0^2}}{{{\rho _0}}}\frac{B}{{2A}}{\rho '^2} + \int_{ - \infty }^t G (t - t')\frac{{\partial \rho '}}{{\partial t}}\mathrm{d}t,
\end{equation}

	\noindent where $\rho '=\rho - \rho _0$ is the density perturbation over the stationary density $\rho _0$, $B/A$ is the nonlinear parameter, $c_0$ is the small amplitude sound speed, and $G(t)$ is the kernel associated with the relaxation mechanism. The first two terms of the right hand side of Eq.~ (\ref{eq:st_cont}) describe the instantaneous response of the medium, where the convolutional third term accounts for the ``memory time" of the relaxing media. Thus, by choosing an adequate time function for the kernel $G(t)$ the model can present an attenuation and dispersion response that fits the experimental data of the heterogeneous media. However, the direct resolution of the constitutive relations Eqs.~(\ref{eq:continuity}-\ref{eq:st_cont}) in this integral form is a complex numerical task due to the convolutional operator. Thus, instead of describe $G(t)$ with a specific time domain waveform, the response of the heterogeneous medium can be alternatively described by a sum of $N$ relaxation processes with exponential time dependence as:
\begin{equation}\label{eq:relax_define}
	\int_{ - \infty }^t G (t - t')\frac{{\partial \rho '}}{{\partial t}}\mathrm{d}t = \sum\limits_{n = 1}^N {{G_n}} * \frac{{\partial \rho '}}{{\partial t}},
\end{equation}

	\noindent with the $n$-th order relaxation kernel expressed as
\begin{equation}\label{eq:define_G}
	{G_n}(t) = {\eta _n}c_0^2\,{\mathrm{e}^{\frac{{ - t}}{{{\tau _n}}}}}H(t),
\end{equation}

	\noindent where $H(t)$ is the Heaviside piecewise function $H(t < 0) = 0$, $H(t > 0) = 1$, $\tau _n$ is the characteristic relaxation time and $\eta _n$ the relaxation parameter for the $n$-th order process. This last dimensionless parameter controls the amount of attenuation and dispersion for each process as $\eta _n = (c_n^2 - c_0^2)/c_0^2$, where $c_n$ is the sound speed in the high frequency limit associated to $n$-th order relaxation process, also known as the speed of sound in the ``frozen" state \cite{Pierce1989}. In order to describe relaxation without the need of including a convolutional operator, we shall define a state variable $S_n$ for each process as
\begin{equation}\label{eq:Sn}
	{S_n} = \frac{1}{\tau _n}{G_n} * \rho '.
\end{equation}

	Thus, using the convolutional property $\frac{\partial }{{\partial t}}\left( {G(t) * \rho '(t)} \right) = \frac{{\partial G(t)}}{{\partial t}} * \rho '(t) = G(t)*\frac{{\partial \rho '(t)}}{{\partial t}}$, the time derivative of the relaxation state variable obeys the following relation for the $n$-th order process:
\begin{equation}\label{eq:dSndt_long}
	\frac{{\partial {S_n}}}{{\partial t}} = \left( { - \frac{1}{{{\tau _n}}}\frac{{{\eta _n}c_0^2}}{{{\tau _n}}}{\mathrm{e}^{ - \frac{t}{{{\tau _n}}}}}H(t) + \frac{{{\eta _n}c_0^2}}{{{\tau _n}}}{\mathrm{e}^{ - \frac{t}{{{\tau _n}}}}}\delta (t)} \right) * \rho ',
\end{equation}

	\noindent where $\delta (t)$ is the Dirac delta function. Using the Eq.~(\ref{eq:Sn}) this relation becomes a simple ordinary differential equation for each process as:
\begin{equation}\label{eq:dSdt}
	\frac{{\partial {S_n}}}{{\partial t}} =  - \frac{1}{{{\tau _n}}}{S_n} + \frac{{{\eta _n}c_0^2}}{{{\tau _n}}}\rho '.
\end{equation}

	Using again convolutional properties, we can substitute Eq.~(\ref{eq:dSdt}) into Eq.~(\ref{eq:relax_define}), and the relaxing nonlinear state Eq.~(\ref{eq:st_cont}) becomes:
\begin{equation}\label{eq:st_final_c0}
	p = c_0^2\rho ' + \frac{{c_0^2}}{{{\rho _0}}}\frac{B}{{2A}}{\rho '^2} - \sum\limits_{n = 1}^N {{S_n}}  + \sum\limits_{n = 1}^N {{\eta _n}c_0^2\rho '} .
\end{equation}

	Moreover, if ``frozen" sound speed for $N$ mechanisms is defined as $c_\infty ^2 = c_0^2\left( {1 + \sum\limits_{n = 1}^N {{\eta _n}} } \right)$, Eq. (\ref{eq:st_final_c0}) leads to:
\begin{equation}\label{eq:st_final_cinf}
	p = c_\infty ^2\rho ' + \frac{{c_0^2}}{{{\rho _0}}}\frac{B}{{2A}}{\rho '^2} - \sum\limits_{n = 1}^N {{S_n}}.
\end{equation}

	Due to the smallness of the relaxation parameter, $\eta _n$, i.e. when weak dispersion is modeled, the sound speed in the high frequency limit reduces to \cite{Naugolnykh1998}:
\begin{equation}\label{eq:cinf}
	{c_\infty } = {c_0}\left( {1 + \sum\limits_{n = 1}^N {\frac{{{\eta _n}}}{2}} } \right).
\end{equation}

	\noindent Note Eq.~(\ref{eq:st_final_cinf}) for a mono-relaxing media is equivalent to that can be found in literature \cite{Rudenko1977}
\begin{equation}\label{eq:st_cinf}
	p = c_\infty ^2\rho ' + \frac{{c_0^2}}{{{\rho _0}}}\frac{B}{{2A}}{\rho '^2} - \int\limits_{ - \infty }^t {\frac{{\eta c_0^2}}{\tau }} {\mathrm{e}^{ - \left( {\frac{{t - t'}}{\tau }} \right)}}\rho '(t')\mathrm{d}t.
\end{equation}

	Thus, the constitutive relations to solve by means of the numerical method in the nonlinear regime are the continuity Eq.~(\ref{eq:continuity}), the motion Eq.~(\ref{eq:momentum}) and the second order fluid state relaxing Eq.~(\ref{eq:st_final_cinf}), where the state variable $S_n$ obeys the relation Eq.~(\ref{eq:dSdt}) for the $n$-th order relaxation process. Although the aim of this work is to model biological media, the generalized formulation presented here can be used to describe the attenuation and hence the dispersion observed in other relaxing media, as the relaxation processes of oxygen and nitrogen molecules in air or the relaxation associated with boric acid and magnesium sulfate in seawater \cite{Pierce1989}.

\subsection{Small amplitude modeling}\label{s:model:linear}
	If small amplitude perturbations are considered, an equivalent derivation of this model can be expressed for multiple relaxation media \cite{Yuan1999}. Thus, for an homogeneous inviscid relaxing fluid, the linearized continuity and motion Eq.~(\ref{eq:continuity}-\ref{eq:momentum}) reduces to
\begin{equation}\label{eq:lin_cont}
	\frac{{\partial \rho }}{{\partial t}} =  - {\rho _0}\nabla  \cdot {\bf{v}}
\end{equation}

	\noindent and
\begin{equation}\label{eq:lin_momen}
	{\rho _0}\frac{{\partial {\bf{v}}}}{{\partial t}} =  - \nabla p;
\end{equation}

	\noindent and linearizing the fluid state Eq.~(\ref{eq:st_final_cinf}) we obtain:
\begin{equation}\label{eq:lin_st}
	\rho ' = \frac{1}{{c_\infty ^2}}\left( {p + \sum\limits_{n = 1}^N {{S_n}} } \right).
\end{equation}

	These equations can be solved directly in this form, however, if expressed in pressure-velocity formulation the density field is no longer necessary and computational effort can be reduced. Thereby, assuming a linear ``instantaneous" compressibility $\kappa _\infty = {\rho _0}c_\infty ^2$, and substituting Eq.~(\ref{eq:lin_st}) into Eq.~(\ref{eq:lin_cont}) yields
\begin{equation}\label{eq:lin_p1}
	\frac{{\partial p}}{{\partial t}} + \sum\limits_{n = 1}^N {\frac{{\partial {S_n}}}{{\partial t}}}  =  - {\kappa _\infty }\nabla  \cdot {\bf{v}}.
\end{equation}

	Then, taking the time derivative of the state variable Eq.~(\ref{eq:dSdt}) we get
\begin{equation}\label{eq:lin_p2}
	\frac{{\partial p}}{{\partial t}} - \sum\limits_{n = 1}^N {\frac{1}{{{\tau _n}}}} {S_n} + \rho '\sum\limits_{n = 1}^N {\frac{{{\eta _n}c_0^2}}{{{\tau _n}}}}  =  - {\kappa _\infty }\nabla  \cdot {\bf{v}}.
\end{equation}

	Finally, substituting again the linearized state Eq. (\ref{eq:lin_st}) and arranging terms the linearized continuity equation leads to
\begin{equation}\label{eq:lin_pconti}
	\frac{{\partial p}}{{\partial t}} + p\sum\limits_{n = 1}^N {\frac{{{\eta _n}c_0^2}}{{{\tau _n}c_\infty ^2}}}  + \sum\limits_{n = 1}^N {\left( {\frac{{{\eta _n}c_0^2}}{{{\tau _n}c_\infty ^2}} - \frac{1}{{{\tau _n}}}} \right){S_n}}  =  - {\kappa _\infty }\nabla  \cdot {\bf{v}}.
\end{equation}

	On the other hand, the state evolution equation can be expressed as a function of the acoustic pressure as
\begin{equation}\label{eq:lin_dSndt}
	\frac{{\partial {S_n}}}{{\partial t}} =  - \frac{1}{{{\tau _n}}}{S_n} + \frac{{{\eta _n}c_0^2}}{{{\tau _n}c_\infty ^2}}\left( {p - \sum\limits_{n = 1}^N {{S_n}} } \right).
\end{equation}

	Thus, the linearized governing Eq.~(\ref{eq:lin_momen}, \ref{eq:lin_pconti}) for a relaxing media are expressed in a pressure-velocity formulation and can be solved together with the coupled state evolution Eq.~(\ref{eq:lin_dSndt}) by means of standard finite differences numerical techniques \cite{Yuan1999}. In this way, lossless linear acoustics equations can be obtained by setting $\eta _n =0$ or in the limit when the relaxation times ${\tau _n} \to \infty $. The relaxation behavior described by this linearized model is achieved too by the formulation described by \citep{Yuan1999}, where the relaxation coefficients $\eta _n$ and the relaxation variable $S_n$ are defined in a different, but analogous way.

\section{Numerical solution by finite-difference time-domain}\label{s:numerical}
	In this section the numerical techniques for solving the complete set of equations (continuity Eq.~(\ref{eq:continuity}), momentum Eq.~(\ref{eq:momentum}), state Eq.~(\ref{eq:st_final_cinf}) and the relaxation Eq.~(\ref{eq:dSdt})) are presented. The numerical method is based on a second order FDTD method where multiple relaxation processes are included in order to: first, modeling physical attenuation and dispersion at the frequencies of interests and second, correct the numerical dispersion and include artificial attenuation to guarantee convergence in nonlinear regime. Moreover the inclusion of relaxation processes in the presented formulation require only one extra field per relaxation process and no memory buffer is needed.

\subsection{Discretization}\label{s:numerical:discr}
\begin{figure}[tb]
	\centering
	\includegraphics{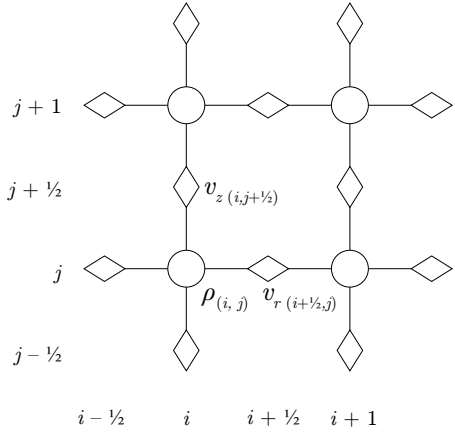}
	\caption{Spatial staggered discretization. The pressure $(p_{i,j}^m)$ and the $n$-th order relaxation process state fields $(S_{n,i,j}^m)$ are evaluated at same discrete location as the density $(\rho _{i,j}^m)$. Particle velocity fields are discretized staggered in both space and time respect to the density, pressure and the $n$-th order state fields.}
	\label{fig1_fdtd_grid}
\end{figure}

	Cylindrical axisymmetric ${\bf{x}} = (r,z)$ coordinate system is considered in this work, however, the method can be derived in other coordinate systems. As in the standard acoustic FDTD method \cite{Botteldooren1996}, the particle velocity fields are discretized staggered in time and space respect to the density and pressure fields. As shown in Fig.~\ref{fig1_fdtd_grid} uniform grid is considered, where $r=i\Delta r$, $z=j\Delta z$, $t=m \Delta t$, with $\Delta r$ and $\Delta z$ as the radial and axial spatial steps, and $\Delta t$ is the temporal step.

	Centered finite differences operators are applied over the partial derivatives of the governing equations. Thus, spatial interpolation is needed over the off-center grid variables in order to fulfill the conservation principles over each discrete cell of the domain \cite{LeVeque1992}. The $r$ component of Eq.~(\ref{eq:momentum}) is expressed in a cylindrical axisymmetric system as
\begin{align}\label{eq:mot_r}
	\frac{{\partial {v_r}}}{{\partial t}} = &- \frac{1}{\rho}\frac{{\partial p}}{{\partial r}} -  {{v_r}\frac{{\partial {v_r}}}{{\partial r}} - {v_z}\frac{{\partial {v_r}}}{{\partial z}}} \\
 \nonumber & + \frac{\eta}{\rho} \left( {\frac{{{\partial ^2}{v_r}}}{{\partial {r^2}}} + \frac{1}{r}\frac{{\partial {v_r}}}{{\partial r}} + \frac{{{\partial ^2}{v_r}}}{{\partial {z^2}}} - \frac{{{v_r}}}{{{r^2}}}} \right) \\
 \nonumber & + \frac{1}{\rho}\left( { \zeta + \frac{1}{3}\eta } \right)\left( {\frac{{{\partial ^2}{v_r}}}{{\partial {r^2}}} + \frac{1}{r}\frac{{\partial {v_r}}}{{\partial r}} + \frac{{{\partial ^2}{v_z}}}{{\partial r\partial z}} - \frac{{{v_r}}}{{{r^2}}}} \right).
\end{align}

	Each term of the above expression is approximated by centered finite differences evaluated at $r = (i+{1\over 2})\cdot\Delta r$, $z = (j+{1\over 2})\cdot\Delta z$, $t = (m+\tfrac{1}{2})\cdot\Delta t$. This equation can be solved obtaining an update equation for ${v_r}_{i+{1\over 2}, j}^{m+{1\over 2}}$. In the same way, the $z$ component of the motion Eq.~(\ref{eq:momentum}) is expressed as
\begin{align}\label{eq:mot_z}
	\frac{{\partial {v_z}}}{{\partial t}} =  &  - \frac{1}{\rho}\frac{{\partial p}}{{\partial z}} -  {{v_r}\frac{{\partial {v_z}}}{{\partial r}} - {v_z}\frac{{\partial {v_z}}}{{\partial z}}} \\
  \nonumber &+ \frac{\eta}{\rho} \left( {\frac{{{\partial ^2}{v_z}}}{{\partial {r^2}}} + \frac{1}{r}\frac{{\partial {v_z}}}{{\partial r}} + \frac{{{\partial ^2}{v_z}}}{{\partial {z^2}}}} \right)\\
  \nonumber &+ \frac{1}{\rho}\left( {\zeta  + \frac{1}{3}\eta } \right)\left( {\frac{{{\partial ^2}{v_r}}}{{\partial z\partial r}} + \frac{1}{r}\frac{{\partial {v_r}}}{{\partial z}} + \frac{{{\partial ^2}{v_z}}}{{\partial {z^2}}}} \right).
\end{align}

	This equation is approximated by centered finite differences and evaluated at $r=i\cdot\Delta r$, $z=(j+{1\over 2})\cdot \Delta z$, $t= m\cdot\Delta t$. An update equation is obtained solving this equation for ${v_z}_{i,j+{1\over 2}}^{m + {1\over 2}}$. Equation~(\ref{eq:continuity}) in cylindrical axisymmetric coordinate system is expressed as
\begin{equation}\label{eq:cont_cyl}
	\frac{{\partial \rho }}{{\partial t}} = -\rho \left( {\frac{{\partial {v_r}}}{{\partial r}} + \frac{{{v_r}}}{r} + \frac{{\partial {v_z}}}{{\partial z}}} \right) - {v_r}\frac{{\partial \rho }}{{\partial r}} - {v_z}\frac{{\partial \rho }}{{\partial z}}.
\end{equation}

	Following the same procedure, each term of the above expression is approximated by centered finite differences and evaluated at $r = i\cdot\Delta r$, $z = j\cdot\Delta z$, $t = (m+{1\over 2})\cdot\Delta t$, and the update equation is obtained solving this expression for $\rho _{i,j}^{m + 1}$. A leap-frog time marching is applied to solve Eq.~(\ref{eq:mot_r}-\ref{eq:cont_cyl}) for each time step until the desired simulation time is reached. Finally, Eq.~(\ref{eq:dSdt}) is locally solved for $m+1$ by and explicit fourth-order Runge-Kutta method and then Eq.~(\ref{eq:st_final_cinf}) is used for update the pressure field.

\subsection{Boundary conditions}\label{s:numerical:bc}

\begin{figure}[b]
	\centering
	\includegraphics{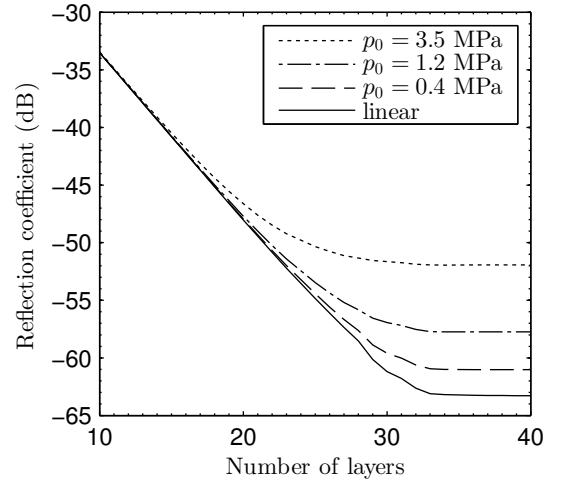}
	\caption{Reflection coefficient of the perfectly matched layer (PML) versus layer thickness for different wave amplitudes.}
	\label{fig2_PML}
\end{figure}

	The staggered grid is terminated on velocity nodes, so the boundary conditions are applied on these external nodes, allowing to prevent the singularity of the cylindrical coordinate system: due to the staggered grid, the only variable discretized at $r=0$ is $v_r$, and axisymmetric condition ${v_r}{|_{r = 0}} = 0$ is applied there. Furthermore, to solve spatial differential operators at boundaries some ``ghost" nodes must be created with the conditions: $v_r(-r)=-{v_r}(r)$, ${v_z}(-r)={v_z}(r)$, $\rho(-r)=\rho(r)$ and $p(-r)= p(r)$.
	
	Perfectly matched layers (PML) \cite{Liu1997} were placed in the limits of the domain ($\pm z$ and $+r$) to avoid spurious reflections from the limits of the integration domain. Inside the PML domains linearized acoustic equations were solved using the complex coordinate screeching formulation \cite{Liu1999}. For a layer of 30 elements and a broadband incident wave with 1 MHz central frequency and non-normal incidence angle, these absorbent boundary conditions have reported a reflection coefficient of $R=-55.2$~dB. However, the performance of the PML is amplitude dependent as long as the nonlinear terms are uncoupled to the PML domains. The amplitude dependence of the reflection coefficient is shown in Fig.~\ref{fig2_PML}, where a PML of 25 layers have reported reflection coefficients $R<-50$~dB for waves in linear regime and highly nonlinear waves including shocks.

\subsection{Minimizing numerical dispersion}\label{s:numerical:stablity}
	The stability for the lossless linear FDTD algorithm follows the Courant-Friedrich-Levy (CFL) condition, that for uniform grid $(\Delta r = \Delta z = \Delta h)$ the maximum duration of the time step is limited by $\Delta t \le \Delta h/{c_0}\sqrt{D}$ where $D$ is the number of dimensions (i.e. $D=2$ in cylindrical axisymmetric coordinate system). That condition essentially states that for a single time step information can not propagate in the numerical grid a distance longer that one cell. However, if relaxation is included numerical instabilities have been observed when ${\tau _f}/2\pi < \Delta t$. Due to this empirical relation, the maximum values for relaxation frequencies are limited too by the chosen spatial discretization by the simple relation ${f_n} < {\sqrt 2}{N_\lambda }{f_0},$ where ${f_n} = 2\pi /{\tau _n}$ is the maximum relaxation frequency for all processes, ${N_\lambda }$ is the number of spatial samples per wavelength and $f_0$ the frequency of the propagating wave.

	On the other hand, nonlinear effects induce the progressively growing of harmonics of the fundamental frequency of the initial wave. The diffusive viscous terms in Eq.~(\ref{eq:momentum}), attenuates the small-amplitude high-spatial frequencies, damping the ``node to node" numerical oscillations and ensuring numerical stability in weakly nonlinear regime. Thus, for a smooth solution the numerical algorithm shows consistency when $\Delta h \to 0$, so if stability is achieved by the CFL condition, the convergence is guaranteed. However, in strongly nonlinear regime, i.e. when sharp waveforms or even shocks are present in the solution, extra numerical techniques must be employed to guarantee convergence. Artificial viscosity can be added when shock waves are present in the solution where a common implementation follows a fourth order spatial filtering \cite{Sparrow1991,Ginter2002}. Thus, the artificial attenuation retrieved by this spatial operator is fourth power of frequency: the low frequency components of the solution remains quasi-undamped, while the higher spatial frequencies are strongly attenuated. In this way, the solution is smoothed and shock thickness depends on the artificial viscosity coefficient.

\begin{figure}[tb]
	\centering
	\includegraphics{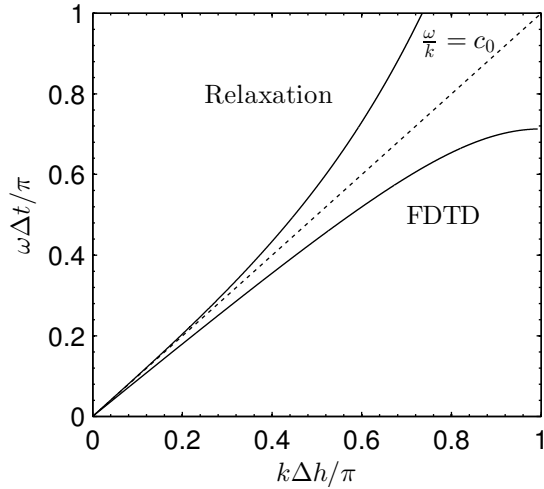}
	\caption{Normalized dispersion relation for of a FDTD lattice of Courant number $S=0.9$ and \emph{anomalous} dispersion relation for a mono relaxing acoustic media. The straight line $c_0=\omega/k$ represents the reference nondispersive case.}
	\label{artificial_dispersion}
\end{figure}

	However, the main drawback for finite difference methods is numerical dispersion, where the analytic dispersion relation can be expressed in 1D as $\sin^2{\frac{k \Delta h}{2}} =\frac{1}{S^2}\sin^2{\frac{\omega \Delta t}{2}}$, with the Courant number $S=c_0\Delta t/\Delta h$. In this way, numerical dispersion reduces phase speed for high frequency components so traveling sharp solutions develop tail oscillations: the low wavenumbers travels fast and left behind high spatial frequencies. In nonlinear regime, is well-known that the combined effects of nonlinearity and strong dispersion can lead rich phenomena, e.g. beatings in the generated harmonics, pulsations on the vertex of a sawtooth wave or soliton formation in strong dispersive media \cite{Rudenko1977}. In this way, the numerical dispersion by discreteness of the FDTD methods couple to the physical nonlinearity can lead to a great variety of non-physical or even unstable solutions. 

	In order to overcome those two limitations, i.e. the generation of harmonics over the discrete limit and the numerical dispersion, we propose the use of artificial relaxation. As Fig.~\ref{artificial_dispersion} shows, physical relaxation processes introduces \emph{anomalous} dispersion, i.e. the phase speed increases in the high frequency regime, opposite to the numerical (lattice) dispersion of the finite differences scheme. Thus, by introducing a collection of relaxation processes and choosing its adequate relaxation parameters the high frequency numerical dispersion can be compensated. As a consequence, introduction of those relaxation processes in the high frequency lead to the inevitable inclusion of artificial attenuation. However, this numerical attenuation is then exploited to limit the growing of higher harmonics in a similar way than artificial viscosity \cite{Sparrow1991}. It is worth noting here that, due to the attenuation using artificial relaxation is, at maximum, only second power of frequency, the low frequency range of the solution is therefore also attenuated. Thus, the proposed method is restricted to lossy media. 
	
	The adequate relaxation parameters that corrects the numerical dispersion have been found by multi-objective optimization techniques, where two cost function are proposed: one for dispersion and other for attenuation. In the first case, the error between the goal (ideal) dispersion relation and the retrieved numerical dispersion corrected by multiple artificial relaxation and is evaluated in the high frequency regime. Finally, the second cost function is the error between the desired (ideal) attenuation and the artificial attenuation evaluated in the low frequency regime.

\section{Validation}\label{s:validation}
\subsection{Single relaxation process}\label{s:val:relax}

\begin{figure}[b]
	\centering
	\includegraphics{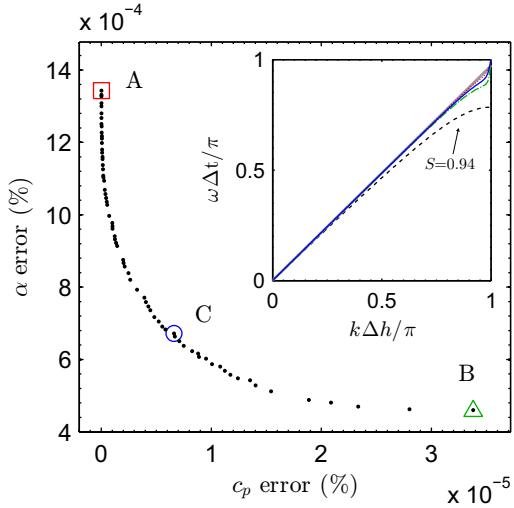}
	\caption{(Color online) Pareto front retrieved by the multi-objective genetic optimization. Square marker (A) is the solution those relaxation parameters minimizes the numerical dispersion. The best fit in the attenuation are the parameters that provided the solution marked by the triangle marker (B). A compromise between attenuation and dispersion is achieved at the individuals around the center of the Pareto front, as shows the sample marked by the circle (C). The inset shows the normalized dispersion relation retrieved by the individuals (A) dotted line, (B) dashed-dotted line, (C) continuous line. Dashed black line shows the numerical dispersion relation of the FDTD method for a Courant number of 0.94 without artificial relaxation.}
	\label{optimized_pareto_single}
\end{figure}

\begin{figure}[b]
	\centering
	\includegraphics{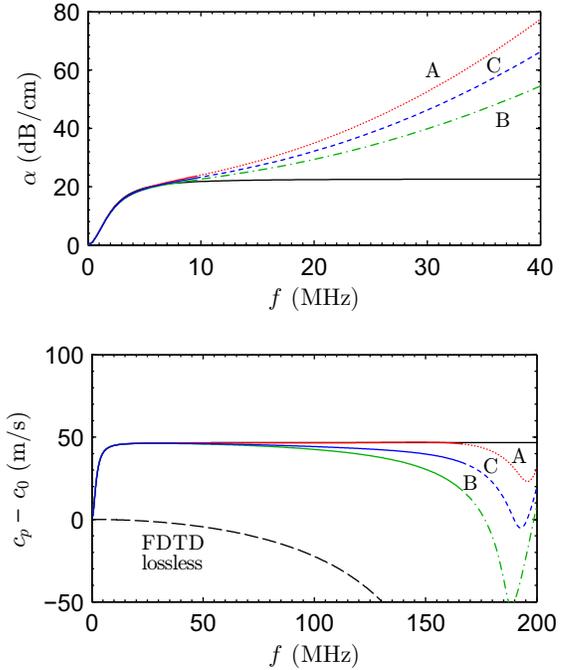}
	\caption{(Color online) Retrieved attenuation (top) and dispersion (bottom) by the inclusion of artificial relaxation for the individuals A (dotted), B (dotted-dashed) and C (dashed) of the individuals marked in Fig.~\ref{optimized_pareto_single}. Continuous lines represent the frequency range included by the optimization, while dashed lines shows the not optimized frequency range. As a consequence of correcting dispersion, attenuation increases in the high frequency range, as shown in the top subplot. This extra attenuation is used as artificial attenuation for numerical-nonlinear stability. }
	\label{optimized_att_and_cphase}
\end{figure}

	A canonical case of a physical single relaxation process is presented. In order to correct numerical dispersion the parameters of two extra artificial relaxation processes have been found using the multi-objective genetic algorithms provided by the optimization toolbox in MATLAB R2014a v8.03. Linear propagation was considered and simulation parameters were set to typical values for water: $c_0=1500$ m/s, $\rho _0=1000$ kg/m$^3$, $B/A=5$, $\eta=8.90 \cdot 10^{-4}$ Pa$\cdot$s. A single physical relaxation process was included, with a characteristic relaxation time of ${\tau _1} = 1/2\pi {f_0}$ and $f_0=2$ MHz, and relaxation modulus of $\eta _1=0.0678$ that leads to a frozen sound speed of $c_\infty=1550$ m/s. In this case, the numerical parameters were set to $\Delta r=\Delta z=1.87\cdot 10^{-7}$ m and $\Delta t=8.65 \cdot 10^{-11}$ s. A plane wave traveling in $+z$ direction was considered.

	Thus, the theoretical attenuation for the relaxation processes and including viscosity can be expressed as
\begin{equation}\label{eq:alpha_t}
	\alpha (\omega ) = \frac{{{\omega ^2}}}{{2{\rho _0}c_0^3}}\left( {\zeta  + \frac{4}{3}\eta } \right) + \sum\limits_{n = 1}^N {\frac{{{\eta _n}}}{{2{c_0}{\tau _n}}}\frac{{{\omega ^2}\tau _n^2}}{{1 + {\omega ^2}\tau _n^2}}}, 
\end{equation}

	\noindent and the theoretical phase velocity can be predicted as \cite{Pierce1989}
\begin{equation}\label{eq:c_t}
	{c_p}(\omega ) = {c_0}\left( {1 + \sum\limits_{n = 1}^N {\frac{{{\eta _n}}}{2}\frac{{{\omega ^2}\tau _n^2}}{{1 + {\omega ^2}\tau _n^2}}} } \right).
\end{equation}

	In order to compute the attenuation and dispersion of the numerical method, simulated pressure was recorded at two locations $z_0$ and $z_1$, and attenuation and phase velocity were estimated from the spectral components over the bandwidth of the input signal. The numerical attenuation was calculated as
\begin{equation}\label{eq:alpha_num}
	\alpha {(\omega)_{fd}} = \frac{{\ln \left( {\left| {P(\omega,{z_1})/P(\omega,{z_0})} \right|} \right)}}{{\left( {{z_1} - {z_0}} \right)}},
\end{equation}

	\noindent where $P(\omega)$ is the Fourier transform of the measured pressure waveforms at points $z_0$ and $z_1$. On the other hand, the phase velocity was computed as
\begin{equation}\label{eq:c_num}
	{c_p}{(\omega)_{fd}} = \frac{{\omega\cdot\left( {{z_1} - {z_0}} \right)}}{{\arg \left( {P(\omega,{z_1})/P(\omega,{z_0})} \right)}},
\end{equation}

	\noindent where correct phase unwrapping is needed in the $\arg$ function.

	In this way, Fig.~\ref{optimized_pareto_single} shows the retrieved Pareto front of the optimization, where 3 different areas can be distinguished. The first area, marked as (A) in Fig.~\ref{optimized_pareto_single}, represents individuals whose numerical dispersion is minimal but attenuation is not optimal. On the other hand, the individuals around area (B) represent a set of relaxation parameters that provides the best agreement between numerical and physical attenuation. A good compromise between both situations can be obtained in the central area of the Pareto front (C), where retrieved attenuation and dispersion in the low frequency band shows good agreement with the physical, and the numerical dispersion has been corrected over a wide frequency range. However, as can be seen in the inset of Fig.~\ref{optimized_pareto_single}, the dispersion relation retrieved by all the cases corrects the FDTD lossless numerical dispersion relation. The phase speed of those tree individuals is shown in Fig.~\ref{optimized_att_and_cphase}~b), where it can be seen that in the frequency range selected for the optimization the numerical phase speed is corrected for all the individuals, where the best fit is obtained for individuals in the Pareto front area A. On the other hand, the inclusion of artificial relaxation leads to an increasing of the attenuation in the high frequency range, as is shown in Fig.~\ref{optimized_att_and_cphase}~a). In this way, as the phase speed error is reduced the effect of artificial attenuation increases. Although this increasing can be seen as a non-desired counterpart, the appearance of this attenuation is useful in order to control the harmonic growing in nonlinear regime in the same way as artificial viscosity spatial operators \cite{Sparrow1991}.

\subsection{Nonlinear steady solution for single relaxation process}\label{s:val:single}

\begin{figure}[tb]
	\centering
    \includegraphics[width=7.5cm]{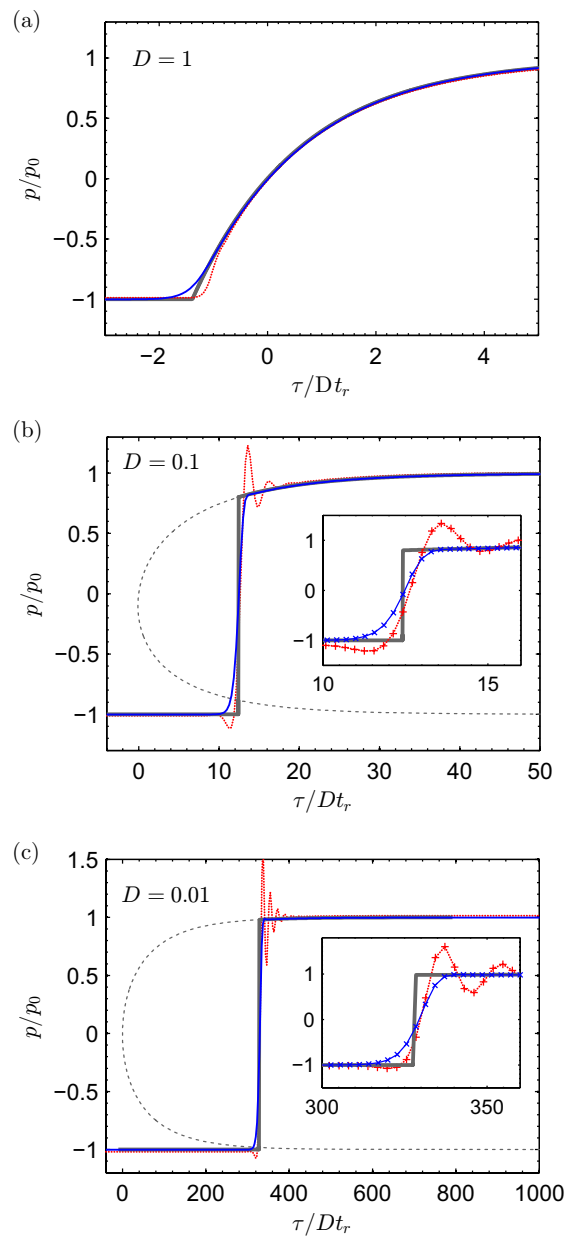}
    \caption{(Color online)~(a) Analytical (thick gray line), numerical using artificial relaxation with corrected dispersion (continuous line) and using artificial viscosity \cite{Sparrow1991} (dotted line) for the nonlinear steady state solution for $D=1$. (b) Nonlinear steady state solution for $D=0.1$. Inset shows detailed shock numerical solution for artificial relaxation ($\times$ markers) and artificial viscosity ($+$ markers). (c) Nonlinear steady state solution for $D$=0.01.}
	\label{Shock_optim_dall}
\end{figure}

	In order to validate the method in the nonlinear regime a full-wave simulation was developed in a mono-relaxing media using above parameters. Thus, the analytical (inverted) solution for the steady solution with $p=-p_0$ for $\tau=-\infty$, $p=p_0$ for $\tau=\infty$ and $p=0$ for $\tau=0$, for the retarded time $\tau=t-z/c_0$ reads \cite{Hamilton1998}
\begin{align}\label{ec:solution_relaxshock}
	\tau  = {\tau _n}\ln \frac{{{{\left( {1 + {p \mathord{\left/
	 {\vphantom {p {{p_0}}}} \right.
	 \kern-\nulldelimiterspace} {{p_0}}}} \right)}^{D - 1}}}}{{{{\left( {1 - {p \mathord{\left/
	 {\vphantom {p {{p_0}}}} \right.
	 \kern-\nulldelimiterspace} {{p_0}}}} \right)}^{D + 1}}}}
\end{align}

	\noindent where $D=\eta _n \rho _0 c_0^2/2\beta p_0$ measures the ratio of relaxation effects to nonlinear effects. For $D>1$, where no shock is present, the solution retrieved by FDTD algorithm shows good agreement with analytical and no artificial attenuation is needed. However, for $D<1$ a discontinuity is present in the solution and convergence is only possible with the inclusion of extra numerical techniques.

	Thus, Fig.~\ref{Shock_optim_dall}~(a-c) shows the analytical and numerical solutions including artificial relaxation and artificial viscosity, where excellent agreement is achieved in all cases. In the case of artificial viscosity, higher harmonics are strongly attenuated and by reducing grid step convergence can be achieved. Due to artificial viscosity operator is essentially a low pass spatial filter, a smoothed version of the shock is achieved. However, the phase speed of the higher spatial frequencies present in the shock is modified due to numerical dispersion, and for $D<<1$ (Fig.~\ref{Shock_optim_dall}~(b, c)) oscillations appears in the tail of the discontinuity, leading to the appearance of non-physical solutions. 
	
	On the other hand, the proposed method of artificial attenuation by relaxation also limits the harmonic growing so a smooth version of the shock appears. Moreover, artificial relaxation also corrects phase velocity so all the spatial frequencies travels at same speed and no oscillatory tail appears. The case of $D=0.01$ is shown in Fig.~\ref{Shock_optim_dall}~(c), where nonlinear effects strongly dominates over attenuation. In this case, tail oscillations provided by artificial viscosity increases in amplitude. In contrast, by including artificial relaxation a smoothed version of the shock is captured and accuracy is maintained.

\section{Results}\label{s:result} 

\subsection{Frequency power law attenuation}\label{s:result:power}

\begin{figure}[t]
	\centering
	\includegraphics{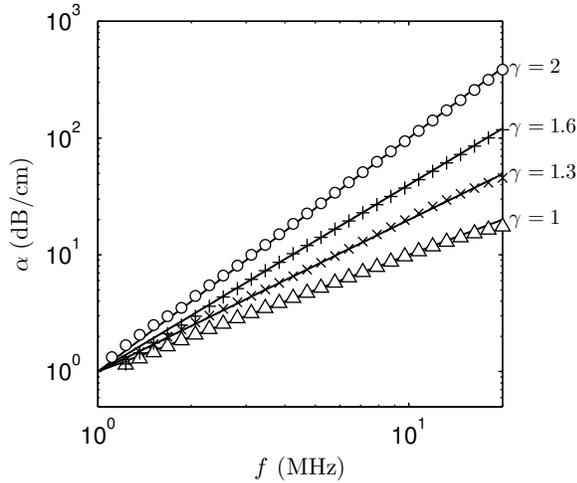}
	\caption{Attenuation retrieved by the numerical algorithm (markers) and target frequency power law attenuation (gray lines). By using the optimization algorithm the relaxation times and modulus were optimized for minimize the relative error between the target power laws of $\gamma=[1, 1.3, 1.6, 2]$ and the attenuation retrieved.}
	\label{fig:attenuation_4}
\end{figure}

	Using methodology described above, the optimal relaxation parameters were obtained in order to fit the multiple-relaxation numerical attenuation to frequency power law attenuation in the form
\begin{align}\label{eq:wgamma}
	\alpha (\omega)=\alpha _0 \omega^\gamma,
\end{align}

	\noindent where $\gamma$ is the power law exponent and $\alpha _0$ the power law coefficient in Np (rad/s)$^{\gamma}$m$^{-1}$. Moreover, the numerical dispersion was corrected by means of artificial relaxation in order to fit the corresponding frequency power law dispersion, where its analytical form satisfying causality can be expressed as \cite{Waters2005}
\begin{align}\label{eq:KKcp}
	\frac{1}{{c_p}\left( \omega  \right)} = {{\frac{1}{{{c(\omega _0)}}} + {\alpha _0}\tan \left( {\frac{{\pi \gamma }}{2}} \right)\left( {{{\left| \omega  \right|}^{\gamma  - 1}} - {{\left| {{\omega _0}} \right|}^{\gamma  - 1}}} \right)}}.
\end{align}

	This expression is valid for $0 <\gamma <3$ with $\gamma \ne 1$, and an alternate equation can be found \cite{Waters2005} in the limit for $\gamma =1$. Here, simulation parameters were $c_0=1500$ m/s, $\rho _0=1000$ kg/m$^3$, $B/A=5$, $\eta=8.90\cdot 10^{-4}$ Pa$\cdot$s, $f_0=1$ MHz, $\Delta r = \Delta z = 1.3 \cdot {10^{ - 5}}$ m, $\Delta t = 5.4 \cdot {10^{ - 9}}$ s; that leads to 26 elements per wavelength and a CFL number of 0.9. Only two independent relaxation processes were employed in this section to obtain the target frequency power laws.

\begin{figure}[t]
	\centering
	\includegraphics{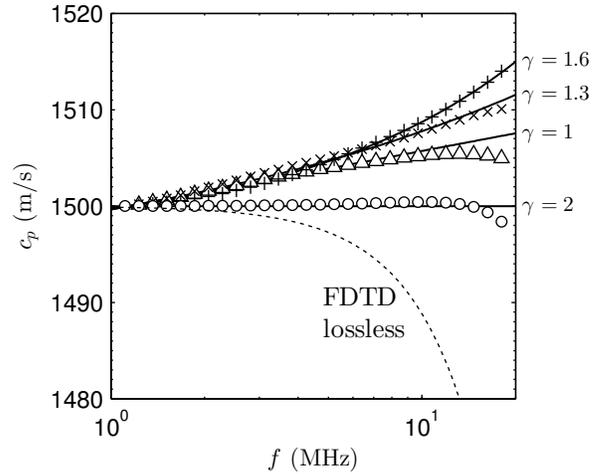}
	\caption{Phase speed retrieved by the numerical algorithm (markers) and target frequency power law attenuation (gray lines) for $\gamma=[1, 1.3, 1.6, 2]$.}
	\label{fig:phase_velocity_4}
\end{figure}

	Following the above procedure, the relaxation times $\tau _n$ and relaxation modulus $\eta _n$ were optimized for different frequency power laws covering the range of that observed in tissues $\gamma=[1, 1.3, 1.6, 2]$. The attenuation coefficient $\alpha _0$ was chosen for each power law to present an attenuation $\alpha=1$ dB/cm/MHz$^\gamma$. The fitting was developed over the typical frequency range for medical ultrasound applications, i. e. 1 to 20 MHz for both attenuation and phase speed. The results for the attenuation and phase speed curves are plotted in Fig.~\ref{fig:attenuation_4} and Fig.~\ref{fig:phase_velocity_4}, where the theoretical and the numerical predictions agree over the frequency range used for the fitting.

\subsection{Fitting attenuation for tissue experimental data}\label{s:validation:exp}

\begin{figure}[b]
	\centering
	\includegraphics{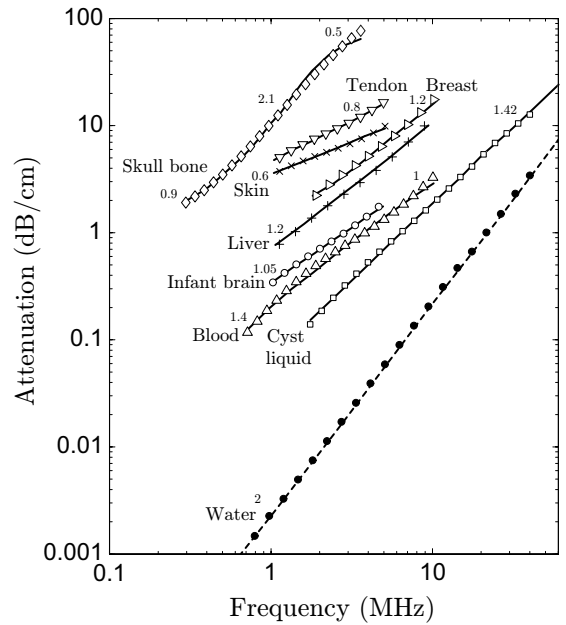}
	\caption{Experimental attenuation data for some tissues adapted from Ref. \citet{Hill2004} (lines), and obtained by the numerical method (markers) by fitting the parameters of 2 relaxation processes. The numbers above the curves show the exponent of the frequency power law $\gamma$ for each frequency region (i.e. the slope of the curve).}
	\label{fig_tissues}
\end{figure}

	Although a frequency power law dependence can describe the ultrasound attenuation over a finite frequency range, the attenuation data of some particular examples shows variation of the exponent over the entire frequency range \cite{Hill2004}. Thus, Fig.~\ref{fig_tissues} shows experimental attenuation data curves for some tissues where the local slope of the power law changes over the measured frequency range. This behavior can be modeled by a sum of relaxation processes by optimizing the relaxation parameters as described above. Thus, the results show that most tissues with locally variable $\gamma$ can be fitted by only a pair of relaxation processes, as the same way that for constant-slope frequency power law attenuation\cite{Cleveland1996}.

	In this way, Table~\ref{table:2:error} shows the error of the numerical attenuation relative to the experimental data. The percent relative error was computed as $\varepsilon  = \frac{{100}}{{{f_2} - {f_1}}}\int_{{f_1}}^{{f_2}} {\frac{{\left| {{\alpha _e}(f) - \alpha (f)} \right|}}{{{\alpha _e}(f)}}df}$, where $\alpha _e(f)$ is the experimental attenuation data, $f_1$ and $f_2$ define the frequency range of the measurement.

\begin{table}[t]
\centering\small
\caption{Error of the optimized attenuation response relative to the experimental data for $N$ total relaxation processes.}
{\begin{tabular}{@{}lccccc@{}}
\toprule
	Tissue 	& 	Power law			&$N=1$					&$N=2$					&$N=3$	&$N=4$		\\
				&	(local slope)		&$\varepsilon(\%)$	&$\varepsilon(\%)$	&$\varepsilon(\%)$	&$\varepsilon(\%)$	\\
\hline
	Skin	 	& 	$f^{0.6}$							& 6.67				& 0.167				& 0.136				& 0.120		\\
	Liver	 	& 	$f^{1.2}$							& 7.62				& 0.517				& 0.404				& 0.165		\\
	Blood 		& 	$f^{1.4}$, $f^{1}$  				& 8.34				& 0.349				& 0.330				& 0.310		\\
	Breast 	& 	 $f^{0.9}$, $f^{1.2}$			& 5.20 			& 0.216				& 0.209				& 0.205		\\
	Skull bone & 	$f^{0.9}$, $f^{2.1}$, $f^{0.5}$	& 10.60			& 10.54				& 8.628				& 5.189\\
\hline
\end{tabular}}
\label{table:2:error}
\end{table}

	As expected, the goodness of fit grows as the number of relaxation processes included increases. However, only two processes are enough to obtain relative errors below 1\% for tissues with $\gamma<2$. In the case of tissues where a local value of $\gamma>2$ has been observed, the fitting procedure fails, like in the skull bone in the 2 MHz range \cite{Hill2004}. The maximum slope achieved by single relaxation and thermo-viscous losses is $\gamma=2$ for any frequency, so a tissue showing that slope cannot be accurately modeled in this frequency region with the method proposed in this work. From another point of view, Eq.~(\ref{eq:KKcp}) states that frequency power law medium with $2<\gamma<3$ presents standard dispersion \cite{Waters2005}, opposite to \emph{anomalous} dispersion for media falling in the range $0<\gamma<2$. Therefore, the dispersion relation of media with $2<\gamma<3$ cannot be modeled by a sum of relaxation processes as long relaxation includes only \emph{anomalous} dispersion.

\begin{table}[b]
\centering\small
	\caption{Variation of sound speed $(\Delta c)$ observed numerically for the modeled tissues by means of two relaxation processes and analytical using the Kramers-Kronig relations.}
	{\begin{tabular}{@{}lcc@{}}
\toprule
		Tissue 	& 	Numerical $\Delta c$		& Analytical $\Delta c$\\
					& 	$(m/s)$					& $(m/s)$\\
\hline
		Skull bone & 	80.737						& 70.720\\
		Skin	 	& 	10.148						& 2.460\\
		Breast 	& 	2.323						& 2.455\\
		Liver	 	& 	3.118						& 2.339 \\
		Blood 		& 	0.865						& 0.907 \\
\hline
	\end{tabular}}\label{table:3:deltac}
\end{table}

	Using Kramers-Kronig relations \cite{ODonnell1981}, the variations of sound speed $\Delta c$ can be predicted by the frequency dependent attenuation. Table \ref{table:3:deltac} shows the variation of sound speed observed in the numerical solution over the fitted frequency range. The magnitude of these variations are of the order of magnitude of those measured experimentally in this frequency range, and the frequency dependence observed for the variation is roughly linear as observed in real tissue\cite{Hill2004}. As expected from the relations between dispersion and absorption \cite{ODonnell1981}, the magnitude of the variation in sound speed increases as the total variation of the absorption increases for a given frequency range.

\subsection{Nonlinear one-dimensional propagation in tissue-like media}\label{s:res:1Dtissue}
\subsubsection{Non-dispersive media}

\begin{figure}[t]
	\centering
	\includegraphics{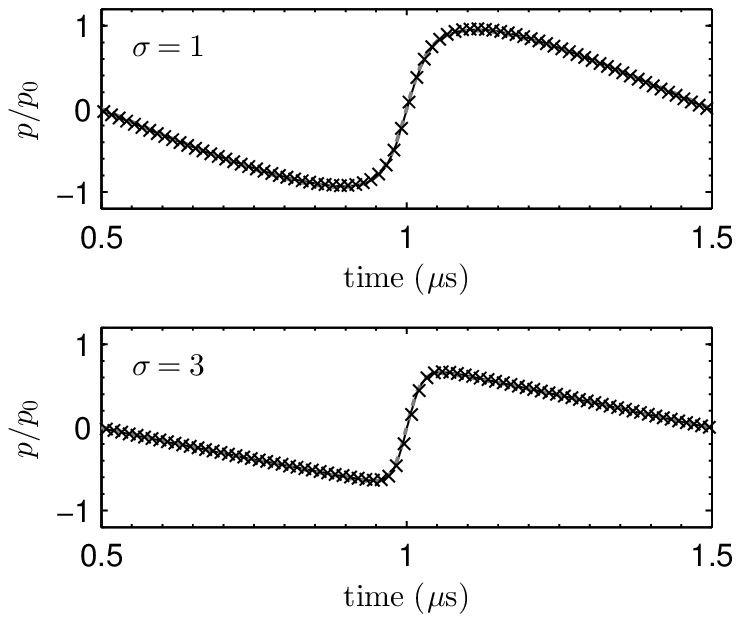}\\
    \includegraphics{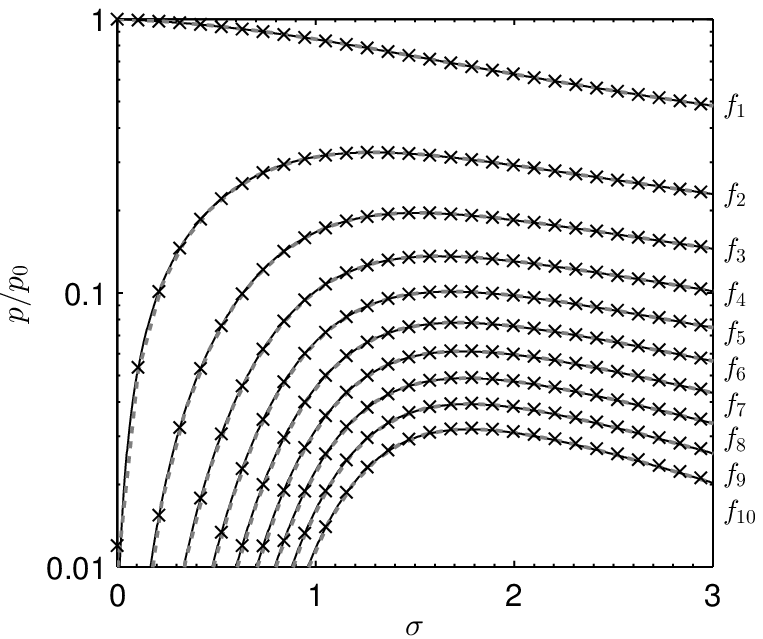}
	\caption{(Top) Waveforms at $\sigma=1$ and $\sigma=3$ for thermo-viscous attenuation ($\gamma=2$). Mendousse analytical solution (black line), $k$-space (gray line) and FDTD numerical solution (markers). (Bottom) Spatial distribution of the first ten harmonics for  Mendousse analytical solution (black line), $k$-space (gray line) and FDTD numerical solution (markers).}
	\label{fig:g2_pt}
    \label{fig:harm_z_g2}
\end{figure}

\begin{figure}[t]
	\centering 
	\includegraphics{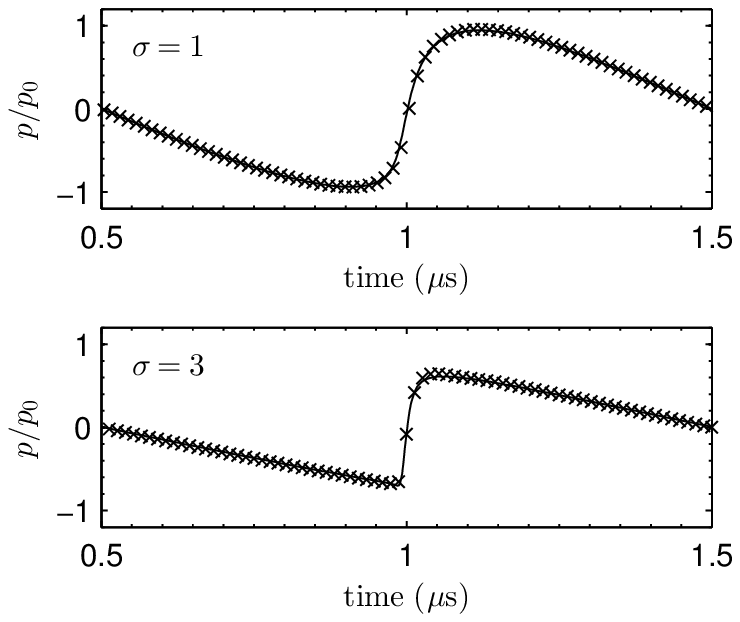}\\
	\includegraphics{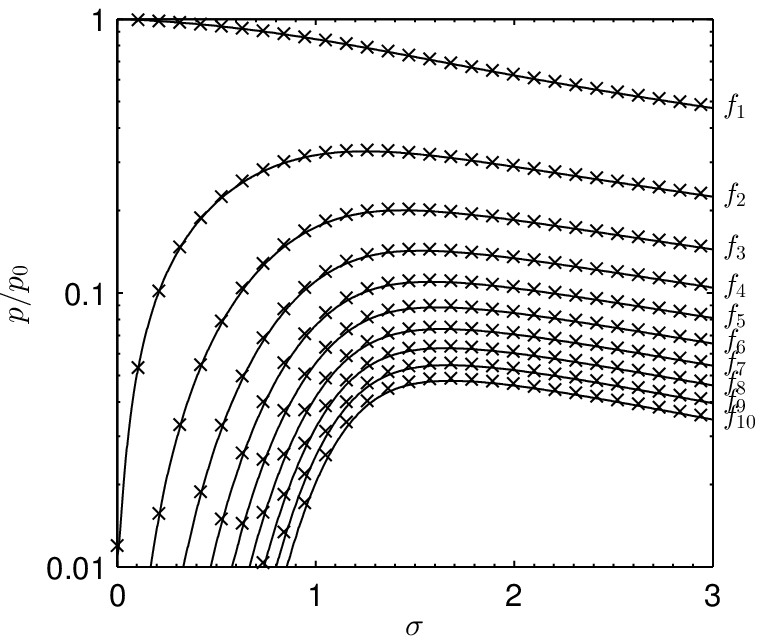}
    \caption{(Top) Waveforms at $\sigma=1$ and $\sigma=3$ in a tissue-like media with frequency power law $(\gamma=1.6)$. $k$-space (gray line) and FDTD numerical solution (markers). (Bottom) Spatial distribution of the first ten harmonics for $k$-space (gray line) and FDTD numerical solution (markers).}
	\label{fig:g16_pt}
 	\label{fig:harm_z_g1_6}
\end{figure}

	In order to study the convergence of the numerical calculations to an analytical solution of the model in the nonlinear regime, a medium with frequency squared dependence attenuation is implemented using the adequate relaxation times and relaxation modulus as explained above. The numerical solution is compared with the analytical solution for a plane wave traveling through a thermo-viscous fluid proposed by Mendousse \cite{Pierce1989}:
\begin{equation}\label{eq:medousse}
	\frac{p}{{{p_0}}} = \frac{{\frac{4}{\Gamma }\sum_{n = 1}^\infty  {{{\left( { - 1} \right)}^{n + 1}}{I_n}\left( {\frac{\Gamma }{2}} \right)n{\mathrm{e}^{ - {n^2}\sigma /\Gamma }}\sin \left( {n\omega t'} \right)} }}{{{I_0}\left( {\frac{\Gamma }{2}} \right) + 2\sum\nolimits_{n = 1}^\infty  {{{\left( { - 1} \right)}^n}{I_n}\left( {\frac{\Gamma }{2}} \right)n{\mathrm{e}^{ - {n^2}\sigma /\Gamma }}\cos \left( {n\omega t'} \right)} }},
\end{equation}

	\noindent where $\Gamma$ is the Gol'dberg number, defined for power law media as $\Gamma= x_a/x_s$, with absorption length $x_a=1/\alpha _0 \omega^\gamma$, shock formation distance $x_s=1/\beta\varepsilon k$ and normalized distance $\sigma=x/x_s$; with the parameter of nonlinearity $\beta=1+B/2A$ and the acoustic Mach number $\varepsilon = v/c_0$, with $v$ is the source particle velocity and $k$ the wavenumber.
    
	Figure~\ref{fig:harm_z_g2}~(top) presents the simulated waveforms at $\sigma=1$ and $\sigma=3$. The wave steepening due to nonlinear processes in the absence of dispersion are well resolved by the numerical method presented here. In order to study the accuracy of the algorithm, the amplitude of the first ten harmonics has been extracted for numerical and analytic solutions and plotted versus $\sigma$ in Fig.~\ref{fig:harm_z_g2}~(bottom). The observed relative error of the computational method decreases due to grid coarsening by a square law (i.e. the numerical scheme is second order accuracy). The magnitude of the error mainly depends on the number of elements per wavelength but, due to not ideal dispersion, also on the traveled propagated distance. Including the correction of dispersion by artificial relaxation, for a path length of 100 $\lambda$, a grid of 26 elements per wavelength was needed to obtain a relative error below 1 \% for the third harmonic. Obviously, the relative error of the first and second harmonics will be always lower, i.e. the fundamental component error was 0.072 \%.

    In addition, the solution was compared also to the obtained by a $k$-space method applied to the constitutive relations, i.e. the $k$-wave algorithm \cite{Treeby2012}. This method was selected due to the low numerical dispersion and the possibility of including frequency power law attenuation. The result of both computational methods and Eq.~(\ref{eq:medousse}) agree over all the spectral components analyzed, showing convergence to the analytic solution.

\subsubsection{Dispersive media}
	In the case of frequency power law attenuation media with $\gamma = (1,2)$ no general analytic solution exist in nonlinear regime for monochromatic progressive waves. Thus, in order to study convergence in this regime, the proposed FDTD solution was compared with the solution obtained by $k$-space methods \cite{Treeby2012}. By using same physical and grid parameters in both methods, the solutions agrees for different power laws. Thus, Fig.~\ref{fig:g16_pt}~(top) shows the good agreement for the waveforms measured at $\sigma = 1$ and $\sigma = 3$ obtained for $\gamma=1.6$, where the characteristic asymmetry effect of media with \emph{anomalous} dispersion (e.g. relaxing, boundary layer effects)\cite{Hamilton1998} is observed: the shock front after the rarefaction phase is followed by a rounded positive compression profile. The spatial distribution for each harmonic is shown in Fig.~\ref{fig:harm_z_g1_6}~(bottom), where it is observed that the proposed FDTD solution with optimized attenuation and dispersion converges to the obtained by pseudo-spectral methods up to ten harmonics. As in the case of frequency squared media, grid refinement numerical tests have reported a second order accuracy of the FDTD method in nonlinear regime.

\subsection{Nonlinear propagation in tissue-like media including diffraction}\label{s:res:3Dtissue}

\subsubsection{Experimental validation}\label{s:res:experiment}

	An experiment was designed to test the validity of the algorithm for intense beams in frequency power law attenuation media. The source was formed by a plane single element piezoceramic crystal (PZ 26, Ferroperm Piezoceramics, Denmark) mounted in a custom designed steel housing and a polymethyl methacrylate (PMMA) focusing lens with aperture $A=50$ mm and radius of curvature $R=50$ mm. The source was driven with a sinusoidal pulse burst of frequency $f_0=1.112$ MHz and $n=50$ cycles using a function generator (14 bits, 100 MS/s, model PXI5412, National Instruments) and a linear RF amplifier (ENI 1040L, 400W, 55dB, ENI, Rochester, NY). The pressure waveforms were acquired with a HNR 500 ${\mathrm{\mu}}$m needle PVDF hydrophone (Onda Corp, CA), and a digitizer (64 MS/s, model PXI5620, National Instruments) was used. A three-axis micropositioning system (OWIS GmbH, Germany) was used to move the hydrophone in three orthogonal directions with an accuracy of 10 ${\mathrm{\mu}}$m. The amplitude frequency response of the hydrophone was compensated in post-processing but not in phase due to the absence of phase calibration for this equipment.

	The source was completely immersed in a castor oil tank ($350 \times 350 \times 350$ mm). We select this frequency power law attenuation media due to the low variability of its acoustic properties along existent literature \cite{Liebler2004,treeby2009}. Using a sound speed inside the bulk of the lens $c_l=2711$ m/s, and a sound speed of the castor oil of $c_0=1480$ m/s (at 26º C room temperature), the effective lens geometrical focal is estimated as $F=R/(1-c_0/c_l)=110.1$ mm, leading to a linear lossless gain of $G=13.4$.  
 
   On the other hand, a nonlinear simulation including diffraction and frequency power law attenuation with same parameters was carried out in a workstation (20 cores Intel Xeon E5-2680 CPU, 2.8GHz with 256 GB RAM). The boundary conditions were implemented for a spherical focused ultrasound source. The castor oil parameters at 26º C room temperature \cite{treeby2009}, were $c_0=1480$ m/s, $\rho _0=961$ kg/m$^3$, $\alpha=0.4$ dB/cm/MHz$^\gamma$, $\gamma=1.69$, $B/A=12.0$. The grid parameters were $\Delta r=\Delta z=29.6$ $\mathrm{\mu}$m and $\Delta t=13.6$ ns, leading to a CFL number $S=0.95$ and $N_{\lambda}=50$ elements per wavelength at fundamental frequency, note that this grid leads to $N_\lambda=16$ for third harmonic.
 
 \begin{figure}[t]
\centering
    \includegraphics{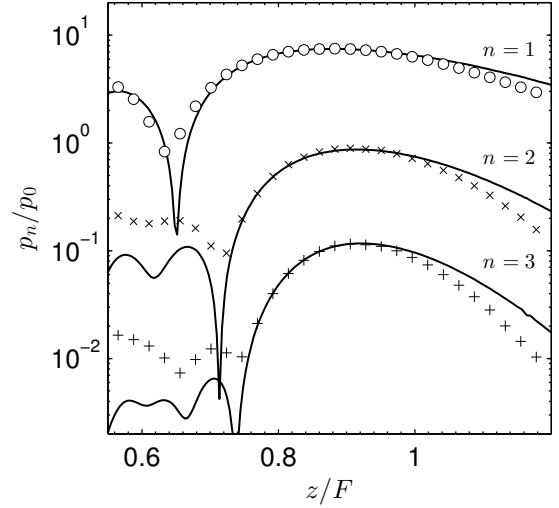}
	\caption{Spatial distribution of the fundamental ($o$), second ($+$) and third ($\times$) obtained by the numerical (continuous line) and experimental methods (markers) for a focused transducer immersed in castor oil.}
	\label{fig:axial_castor}
\end{figure}

   The balance between nonlinear effects and power law attenuation can be estimated by using the Gol'dberg ratio, $\Gamma=x_a/x_s$ where $x_s$ is the shock formation distance and $x_a$ the media attenuation characteristic length. Thus, the amplitude of the source were selected to obtain a Gol'dberg ratio of $\Gamma=0.25$ in order to let the frequency power law attenuation effects slightly dominate over nonlinear effects. On the other hand, the ratio between diffraction effects and nonlinear effects can be described by the so called Khokhlov number \cite{Naugolnykh1998} as $N_\mathrm{K}=x_s/x_d$, where $x_d= k a^2/2$ is the diffraction length and $a$ the source radius. For the proposed test a Khokhlov number of $N=0.5$ was selected to let the nonlinear effect slightly dominate over diffraction effects. The selected excitation pressure amplitude was $p_0=87.7$ kPa.

	The results are summarized in Fig.~\ref{fig:axial_castor}, where axial pressure distribution for the fundamental, second and third harmonic are presented. A good agreement is found between simulations and the experimental test. Only far to the focal point the amplitude there exist differences between computations and experiments, that can be caused by nonuniform vibration of the source \cite{canney2008}, boundary effects of the PMMA lens, or miss-alignment of the source axis and micro-positioning system orthogonal directions along the 100 mm axial measurement. 
    
    The maximum amplitudes of the first harmonic were $p_{e1}=0.6539$ MPa for the experiment and $p_{n1}=0.6576$ for the numeric. The second harmonic peak pressure was $p_{e2}=78.368$ kPa and $p_{n2}=76.272$ kPa, and the third harmonic peak pressure $p_{e3}=10.146$ and $p_{n3}=10.252$ kPa for the experimental and numeric respectively. The relative errors between numerical and experimental results are $0.56$ \% for the fundamental frequency, $2.67$ \% for the second harmonic and $1.04$ \% for the third. No error estimation was done for the peak pressure due to the absence of a phase calibration of the hydrophone.

\subsubsection{Highly focused beam}\label{s:res:hifu}

\begin{figure}[t]
\centering
    \includegraphics{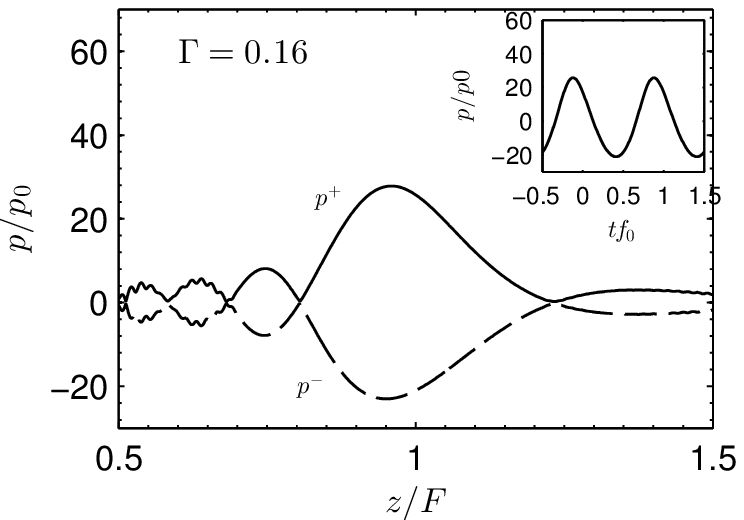}\\
    \includegraphics{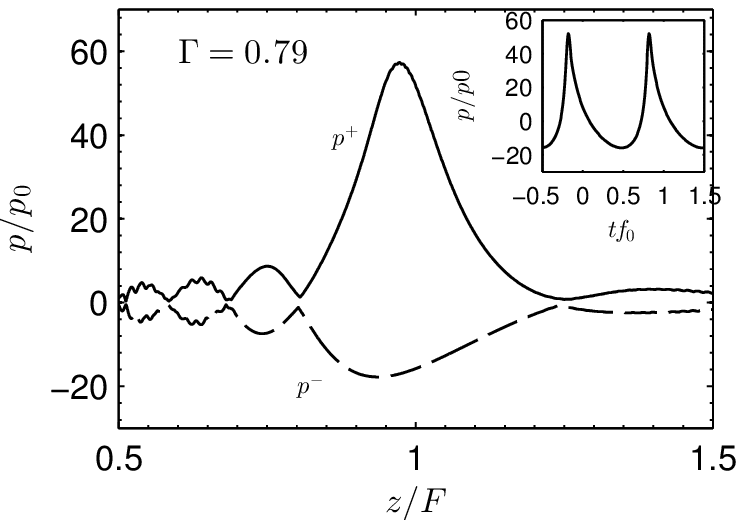}
	\caption{Axial spatial distribution of the peak compression ($p^+$) and minimum rarefaction ($p^-$) pressure for a focused beam propagating through a liver tissue layer. (Top) weakly nonlinear propagation ($\Gamma=0.16$) and (bottom) strong nonlinear effects ($\Gamma=0.79$). The insets show the waveforms recorded at the geometrical focal.}
	\label{fig:cut_insets}
\end{figure}

\begin{figure}[t]
\centering
    \includegraphics{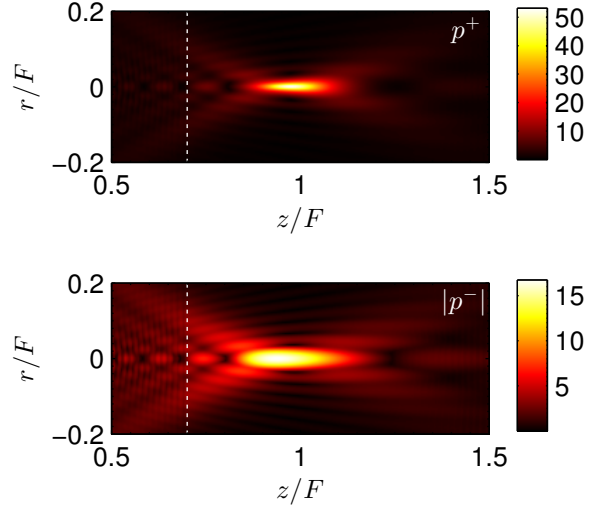}
	\caption{(Color online) Spatial distribution of the peak compression ($p^+$) and minimum rarefaction ($p^-$) pressure for a focused beam propagating through a liver tissue layer. (boundary marked with dashed line) for $\Gamma=0.79$. Colorbars are in $|p|/p_0$ units.}
	\label{fig:mapp_10}
\end{figure}

	In order to test the algorithm in the very high nonlinear regime with realistic tissue parameters a focused bowl of geometrical focal $F=50$ mm and aperture $A=50$ mm, driven at $f_0=1$ MHz was numerically tested. These parameters leads to a source gain $G=26.5$ and a $f$-number$=1$, showing that source is beyond the paraxial limit. The media consist in two layers. The first layer, where the source was located, was water at 20º C with parameters $c_1=1482$ m/s, $\rho _1=1000$ kg/m$^3$, $B/A_1=5$, $\alpha_1=2.17 \times 10^{-3}$ dB/cm MHz$^{\gamma}$, $\gamma _1=2$. At a distance $z_l=0.7/F=35$ mm, a layer of human liver tissue was placed, therefore the focal spot is located inside tissue at a depth of 15 mm. Liver tissue parameters \cite{Hill2004} were $c_2=1597$, $\rho_2=1050$ kg/m$^3$, $B/A_2=7.9$, $\alpha_2=.75$ dB/cm MHz$^{\gamma}$, $\gamma _2=1.5$. In this case, the numerical grid parameters were $\Delta r=\Delta z=29.6$ $\mathrm{\mu}$m and $\Delta t=11.9$ ns, leading to a CFL number $S=0.9$ and $N_{\lambda}=50$ elements per wavelength at fundamental frequency ($N_\lambda=16$ for third harmonic).

	Figure~\ref{fig:cut_insets} shows the spatial peak pressure profiles for different excitation amplitudes. In Fig.~\ref{fig:cut_insets}~(top) the pressure of the source was $p_0=0.18$ MPa, while in the bottom figure was increased to $p_0=0.94$ MPa. Thus, for the selected parameters, Fig.~\ref{fig:cut_insets}~(top) present results for $\Gamma=0.16$ and $N_\mathrm{K}=0.62$, so the attenuation effects dominates over nonlinear effects and nonlinearity slightly dominates over diffraction effects. In this way, low asymmetry is observed between the positive compression peak, $p^+$, and the minimum rarefaction pressure distribution, $p^-$. The calculated waveform at $z=F$, shown in the inset of Fig.~\ref{fig:cut_insets}~(top), is weakly distorted. However, there exist differences between its normalized peak amplitude $p^+/p_0=25.7$ and $p^-/p_0=21.05$, and the source characteristic linear gain, $G=26.5$. They are caused, in one hand by the attenuation effects, where the value of the lossy linear gain observed was $G_\alpha=p^+/p_0=p^-/p_0=23.1$ measured at $z=F$, i.e. the amplitude at the focal was reduced to 87.2\% of the lossless amplitude. On the other hand, the differences due to the asymmetry between compression and rarefaction cycles are caused by the combined effect of nonlinearity and focusing.

	If source amplitude is increased to $p_0=0.94$ MPa, as shown in Fig.~\ref{fig:cut_insets}~(bottom), the ratio between attenuation and nonlinear effects is increased to $\Gamma=0.79$. In this regime, nonlinear effects are almost of the same order of attenuation effects. On the other hand, increasing source amplitude while keeping same transducer parameters implies also the Khokhlov number changes to $N_\mathrm{K}=0.12$, so the nonlinearity clearly dominates over diffraction effects. In this regime, highly asymmetric pressure distribution is observed, where the values at focal point are $p^+=49.05$ MPa and $p^-=-14.91$ MPa.

    Other typical nonlinear phenomena characteristic of high intensity focused sources can also be observed: formation of sharp shock front and its corresponding harmonic generation, or, as Fig.~\ref{fig:mapp_10} shows, narrowing of the beam for $p^+$ and broadening for $p^-$ pressure distributions. In addition, nonlinear focal shift, i.e. displacement of the peak pressure relative to the position of the linear peak pressure can be also predicted for tissue propagation. In the case of $\Gamma=0.79$ it was observed a nonlinear focal shift $\Delta F^+=+1.05$ mm and $\Delta F^-=-1.03$ mm for the $p^+$ and $p^-$ pressure distribution respectively.

\section{Summary}\label{s:summary}
	A general model based on the full constitutive relations of nonlinear acoustics in relaxing media have been presented in a time-domain formulation which does not require convolutional operators. A numerical solution by means of finite-differences in time-domain have been obtained, showing that the theoretical attenuation and dispersion due to relaxation processes can be achieved by the numerical method with accuracy. These results can be also used to model typical relaxation process of other relaxing media (e. g. the processes observed in air, associated to the molecules of oxygen and nitrogen, or in seawater, associated to the relaxation of boric acid and magnesium sulfate). 

	Moreover, a method for modeling frequency power law attenuation by means of multiple relaxation has been implemented in the constitutive relations. The proposed method can describe local variations of the exponent of the frequency power law, so an arbitrary attenuation curve in the range $0<\gamma<2$ can be modeled by means of the proper optimization of the relaxation coefficients. This feature of the presented method is an advantage when compared with most fractional derivatives methods, where the attenuation follows an exact but unique frequency power law over the entire frequency range. A broad range of human tissues have been modeled and the goodness of the fit using from two to four relaxation processes has been discussed. 

	Furthermore, a computational technique that exploits the \emph{anomalous} dispersion of the relaxation processes is employed to mitigate the numerical dispersion of the finite-differences scheme. Thus, while phase speed is corrected by including artificial relaxation processes, its corresponding artificial attenuation is used to improve stability in the nonlinear regime. In this way, smooth and stable versions of shock waves have been obtained and compared with its analytic solution. Furthermore, the validity of the algorithm including diffraction have been tested with experimental measurements of a focused beam in castor oil.
    
	Due to the model is developed from the constitutive relations for nonlinear acoustics, most wave phenomena is captured. As a difference from the one-way models the proposed model implicitly includes multiple wave direction, and, due to the Lagrangian density of acoustic energy is implicitly included in the computation, multiple scattering and strong resonance effects can be accurately described. Moreover, unlike KZK and other parabolic approximations, the proposed model captures the diffraction exactly, so for simulation of acoustic beams the field is not approximated only to the beam axis, but also in the near field and far to the beam axis and thus high focused devices can be simulated. 
    
    The code has shown to be particularly appropriate if the problem to simulate presents axisymmetry, because the constitutive relations for nonlinear acoustics are solved in a computational 2D domain, while standard $k$-space methods need to employ full 3D domains due to the poor convergence of the Fourier series at discontinuities ($r=0$). This is the case, for example, of the focused ultrasound transducer simulated in Sec.~\ref{s:res:hifu}, where a full 3D solution will require huge computational resources and calculation times. Finally, due to the particle velocity vector and the acoustic density fields are solved implicitly by the code, this information can be used to estimate other relevant magnitudes as the full nonlinear intensity vector, the nonlinear acoustic radiation forces in these absorbing media or the acoustic streaming generated in frequency power law attenuation fluids. 
 
\section*{ACKNOWLEDGMENTS}
The authors acknowledge financial support from the FPI program of the Universitat Polit\`ecnica de Val\`encia.


%




\end{document}